\documentclass[journal=jpc,manuscript=article]{achemso}

\usepackage[version=3]{mhchem} 
\usepackage{graphicx}
\usepackage{amsmath}
\usepackage{amssymb}
\usepackage{amsfonts}
\usepackage{soul}
\usepackage{dcolumn}
\usepackage{bm}
\usepackage{hyperref}
\usepackage[mathlines]{lineno}
\usepackage{xcolor}
\usepackage{titlecaps}
\definecolor{sph}{rgb}{0.0588, 0.3216, 0.7294} 
\definecolor{ppk}{rgb}{1.0, 0.4549, 0.0902}

\newcommand{\nc}{\newcommand}
\nc{\nn}{\nonumber}
\nc{\txt}{\textrm}
\nc{\txtsup}{\textsuperscript}
\nc{\txtsub}{\textsubscript}
\nc{\calL}{\mathcal{L}}
\nc{\U}{\mathcal{U}}
\nc{\T}{\mathcal{T}}
\nc{\E}{\mathcal{E}}
\nc{\XYZ}[1]{\textcolor{black}{#1}}
\nc{\XXZ}[1]{\textcolor{black}{#1}}
\nc{\XZZ}[1]{\textcolor{black}{#1}}
\newcommand{\onlinecite}[1]{\hspace{-1 ex} \nocite{#1}\citenum{#1}} 

\author{Subhajit Sarkar}
\email{subhajit@post.bgu.ac.il}
\affiliation{Department of Chemistry, Ben-Gurion University of the Negev, Beer Sheva 84105, Israel,}
\affiliation{School of Electrical and Computer Engineering, Ben-Gurion University of the Negev, Beer Sheva 84105, Israel.}

\author{Yonatan Dubi}
\email{jdubi@bgu.ac.il}
\affiliation{Department of Chemistry, Ben-Gurion University of the Negev, Beer Sheva 84105, Israel,
}
\affiliation{Ilse Katz Center for Nanoscale Science and Technology, Ben-Gurion University of the Negev, Beer Sheva 84105, Israel.
}

\date{\today}

\title{Emergence and Dynamical Stability of Charge Time-Crystal in a Current-Carrying Quantum Dot Simulator}

\keywords{ Quantum Dots Arrays, Discrete Time-Crystals, Quantum Simulation,  Dissipative Quantum Systems, Dynamical Stability}
\makeatletter
\newcommand*{\forcekeywords}{
  \acs@keywords@print
  \let\acs@keywords@print\relax
}

\begin{document}
\forcekeywords


\begin{center}
\date{\today}
\end{center}
\begin{abstract}

Periodically-driven open quantum systems that never thermalize exhibit a discrete time-crystal behavior, a non-equilibrium quantum phenomena that has shown promise in quantum information processing applications. Measurements of time-crystallinity are currently limited to (magneto-) optical experiments in atom-cavity systems and spin-systems making it an indirect measurement. We theoretically show that time-crystallinity can be measured directly in the charge-current from a spin-less Hubbard ladder, which can be simulated on
quantum-dot array. We demonstrate that one can dynamically tune the system out and then back into the time-crystal phase, proving its robustness against external forcings. These findings motivate further theoretical and experimental efforts to simulate the time-crystal phenomena in current-carrying nano-scale systems.
\end{abstract}




Schr\"{o}dinger's equation, which governs quantum physics, is time-translation invariant. Despite that, time-translation symmetry breaking has been shown to be possible in quantum systems under specific circumstances, leading to the so-called discrete time-crystal (DTC) behavior in numerous systems including, atom-cavity systems, spin-ensembles, solid-state quantum-dot (QD) arrays, and quantum processors \cite{Choi2017, Zhang2017,Kyprianidis1192, pal_prl_cluster, magnon_space_time_crystals_prl, smits_prl_superfluid_QG, Zhang2017, Dogra_science.aaw4465, Autti2021, Economu_PhysRevB_2019, Qiao2021, vandyke2020protecting, randall2021many, mi2021time}.
Periodically driven (Floquet) closed quantum systems that never reach a thermodynamic equilibrium can indeed exhibit a DTC behavior \cite{  Sacha_2017, Khemani_PRL_DTC,Else_PRL_Floquet_DTC, Surace_prb_DTC, khemani2019brief, Wilczek_prl_quantum_tc, Choi2017, Zhang2017, randall2021many, mi2021time}, manifested in the sub-harmonic oscillations of the order parameter. They typically rely on disorder and localization to avoid reaching a stationary state of infinite temperature that opposes time-crystalline order \cite{Yao_Nayak_TC_physics_today, bukov_polkovnikov_flloquet_2015,Khemani_PRL_DTC, Moessner2017}. Alternatively, dissipation and fluctuations induced by controlled coupling to suitable environments stabilize DTC in systems without disorder and localization \cite{RieraCampeny2020timecrystallinityin, Buca_nat_comm2019, diss_Flq_DTC_PRL_Ikeda, BTC_PhysRevLett_Fazio,carollo2021exact, Gong_PhysRevLett, Marino_Demler_NJP_2019, Gambetta_PhysRevLett_metaDTC, Buca_PhysRevLettQG_2019, Prosen_SciPostPhys_2020, Tindall_2020, chinzei2021criticality, Gunawardana2021DynamicalLI, buca2021algebraic,Buca_prb_2020, Buca_PhysRevLettQG_2019, Prosen_SciPostPhys_2020, Tindall_2020}, which was confirmed in the recent observation of dissipative DTC in the atom-cavity system \cite{Dissipative_DTC_PhysRevLett_atom_cavity}. 
Observations of DTCs in both closed and open Floquet quantum systems have been limited to optical detection via spin-dependent fluorescence from spin-full systems, e.g., non-interacting spin ensembles, spin-Boson, Ising and XXZ systems \cite{Choi2017, Zhang2017, Kyprianidis1192, mi2021time, Choi2017, Zhang2017, Kyprianidis1192, pal_prl_cluster, magnon_space_time_crystals_prl, smits_prl_superfluid_QG, Qiao2021, vandyke2020protecting, Dissipative_DTC_PhysRevLett_atom_cavity, randall2021many}. To this end, the atom-cavity systems and optical lattice set-ups have been dominant platforms for achieving DTC due to their high degree of tunability in terms of dissipation engineering.

Although DTC phenomena is interesting in its own right, applications of DTCs are only beginning to emerge.  Advanced fabrication and control of coupled quantum dot (QD) arrays \cite{Takumi_APL_2018, Mukhopadhyay_APL_2018, Mills2019, Sigillito_PhysRevApplied2019, Qiao_PhysRevX_2020} led to the theoretical proposal of observing \cite{Economu_PhysRevB_2019} and subsequent experimental realization of a DTC in nano-scale system, attempting stable quantum information processing using DTC as a mechanism \cite{Qiao2021, vandyke2020protecting, EstarellasSci_Adv8892}. Gate defined QD array, a.k.a, Quantum-Dot simulator has been used as a platform to investigate and realize multitude of phenomena that include Fermi-Hubbard model and Mott transition \cite{Hensgens2017}, Heisenberg models \cite{Qiao_PhysRevX_2020, Qiao2021, Diepen_PhysRevX} and phonon-induced pairing \cite{Bhattacharya_nano_lett}. 
However, it seems that the possibility of detecting DTC behavior with charge transport measurements has not been explored. This is somewhat surprising, since charge transport measurements, e.g., through quantum dot (QD) arrays, allowed for a large range of discoveries in quantum physics, from Coulomb blockade to Kondo physics, Superconductivity, quantum-point-contact universal conductance, and quantum computing \cite{Field_PhysRevLett, Byrnes_PhysRevB2008, Yang_PhysRevB2011, Barthelemy_2013, Deshpande2010, Loss_DiVincenzo_PhysRevA}. A possible reason for that is that it is a-priori unknown if DTCs survive or die when there is a charge flow through the system, and previous studies of DTCs (in spin-systems) suggest that fast decay will overshadow any signature of the DTC \cite{Lazarides_Moessner_2017, Marino_Demler_NJP_2019}.
 
Here we investigate if a dissipative-DTC can be found in a current-carrying spin-less system and how it can be observed in a transport experiment. In a spin-less system, the charge density is the primary observable. Therefore, it suits best to investigate charge transport in a dissipative-DTC state. We suggest a route for directly measuring dissipative-DTC behavior in transport measurements in a periodically-driven two-leg spin-less Hubbard ladder, which a quantum dot array can simulate \cite{Bhattacharya_nano_lett, Lawrie_nanolett, Jang_nanolett, Riggelen_APL}. Specifically, we discuss system parameters that are relevant to QD arrays \cite{Bhattacharya_nano_lett, Lawrie_nanolett, Jang_nanolett}, where periodic-driving can be achieved in an electrically tunable set-up \cite{Takumi_APL_2018, Mukhopadhyay_APL_2018, Mills2019, Sigillito_PhysRevApplied2019, Qiao_PhysRevX_2020}. The main advantage of Quantum-Dot simulators over the optical-lattice ones is the relative ease of transport measurements in the former.

Our main results are as follows: (i) We theoretically show that a stable and a meta-stable dissipative DTC \footnote{We use the term discrete time crystal also to describe discrete-time quasi-crystals \cite{DTQC_PhysRevB_Sacha}, where the ratio between the driving and response frequencies in an irrational number.} can be observed via measurement of the charge-current, and identify the conditions necessary for this to happen. We show that The DTC manifests itself in a sharp peak of the discrete Fourier transform (DFT) of the charge-current, which is locked at the DTC sub-harmonic frequency, even in the meta-stable regime where the oscillation develops a decaying envelope. To this end we clarify that the notion of meta-stability we use pertains to the finite lifetime of the DTC. (ii) We show that a reversible transition from a DTC to a normal phase can be driven by dynamically tuning the system parameters. Since the system is current-carrying (and hence is an {\sl open quantum system}), this dynamical transition is distinct from the existing notion of dynamical transition between DTC phases proposed in closed (dissipation-less) quantum systems \cite{Yang_PhysRevLett_dynamical_transition_DTCS,kosior2018dynamical}.

Our theory reveals the fundamental origin of the DTC behavior in a spin-less nanoscale system. We show that the two-leg spin-less Hubbard ladder, that can be emulated through a QD array \cite{Riggelen_APL}, exhibits a DTC behavior because of the hidden SU(2) symmetry, arising from its ladder structure. Under proper engineering of dissipation the hidden SU(2) symmetry ensures the emergence of a weak-local Floquet-Dynamical symmetry (FDS) \cite{sarkar2021signatures} in the system. Indeed, as detailed in Ref.~\onlinecite{Buca_nat_comm2019}, dissipation (in the form of on-site dephasing) is necessary for inducing the DTC behavior, because it allows for coupling of different charge sectors of the Floquet Hamiltonian which, in absence of dephasing, would otherwise remain decoupled. This weak-local FDS provides both the mechanism for the existence and the necessary decay for the measurement of DTC behavior \cite{Volovik2013}. The mechanism of weak-local FDS in our model is fundamentally different from the existing notion of FDS in spin-full systems which is based on an effective Zeeman effect \cite{diss_Flq_DTC_PRL_Ikeda, Buca_nat_comm2019, Buca_prb_2020, Buca_PhysRevLettQG_2019, Prosen_SciPostPhys_2020, Tindall_2020}. Understanding the underlying mathematical structure would motivate further search of dissipative DTC in various other nanoscale systems \cite{Petiziol2021, Hussain_jacs_2018, Chiesa_jpclett2020}. 

\begin{figure}
 \centering
 \includegraphics[keepaspectratio=true,scale=0.4]{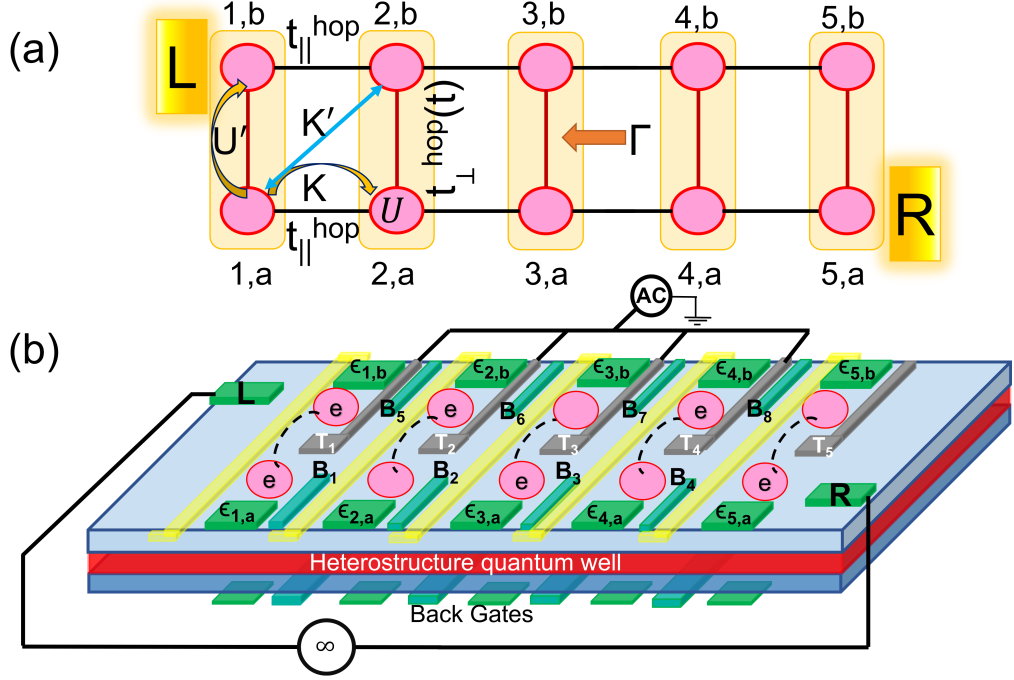}
\caption{(a) Schematic representation of the two-leg ladder Hamiltonian with legs $\alpha = a$ and $b$. $t_{||}^{hop}$ is the nearest neighbor hopping strength in each leg, $K$ is the nearest neighbor interaction, $U$ is the onsite interaction for each leg; $U^{\prime}$ is the inter-chain nearest-neighbor interaction, and $K^{\prime}$ is the next nearest-neighbor interaction. $t_{\perp}^{hop}(t)$ is the time-dependent inter-ladder hopping which is periodically driven. (b)  Schematic picture of a possible (experimental) set-up of an array of 10 gate controlled QDs: (green) $\epsilon_{j,\alpha}$'s are the gates that control the onsite energies. $B_{j}$'s are the tunnel barrier that provide a control on $t_{||}^{hop}$, $T_{j}$'s control the tunnel barrier which are periodically driven with an AC of frequency $\omega$ that provide $t_{\perp}^{hop} (t)$. Yellow channels represent suitably placed quantum-point contacts that provide a possible dephasing mechanism \cite{Granger2012}.  $\sqrt{\Gamma}$ is the strength of the dephasing, which is further indicated by a rounded box in (a). $L$ and $R$ represent the left and right electrodes, respectively where suitable bias can be added to drive electric current through the system. }
 \label{fig_set-up}
 \end{figure}

We consider a periodically driven ladder of spin-less Fermions, see Fig. \ref{fig_set-up} \XZZ{(schematic set-up is motivated by Ref. \onlinecite{benenti2019principles})}, comprising of $2N$ inter-connected sites with a Hamiltonian, $\mathcal{H}(t) = \mathcal{H}_{0} + \mathcal{H}_{hop}^{||} +\mathcal{H}_{int} + \mathcal{H}_{hop}^{\perp}(t)$ where \cite{diss_Flq_DTC_PRL_Ikeda, Hensgens2017}
\begin{eqnarray}\label{Eq:ham}
\mathcal{H}_{0} &=& \sum_{\substack{j =1, \\ \alpha = a,b}}^{N-1} \epsilon_{\alpha} n_{j, \alpha},~~ \mathcal{H}_{hop}^{||}= -t_{||}^{hop} \sum_{\substack{j =1, \\ \alpha = a,b}}^{N-1}  (c_{j,\alpha}^{\dagger} c_{j+1,\alpha} + c_{j+1,\alpha}^{\dagger} c_{j,\alpha}) \nonumber \\ \mathcal{H}_{int} &=&  \sum_{j=1}^{N}\sum_{\alpha =a,b}\frac{U}{2}n_{j, \alpha} n_{j, \alpha}  + \sum_{j=1}^{N-1}\sum_{\alpha=a,b}\frac{K}{2} n_{j, \alpha} n_{j+1, \alpha} + \sum_{j=1}^{N-1} \frac{K^{\prime}}{2} (n_{j,a} n_{j+1,b}+n_{j+1,a} n_{j,b}) \nn \\&+& \sum_{j=1}^{N} U^{\prime} n_{j,a} n_{j,b}, \nn \\
 \mathcal{H}_{hop}^{\perp}(t) &=& -\sum_{j=1}^{N}\left[ t_{\perp}^{hop}(t) c_{j,b}^{\dagger} c_{j,a} + [t_{\perp}^{hop}(t)]^{*} c_{j,a}^{\dagger} c_{j,b}  \right],
\end{eqnarray}
where $\epsilon_{\alpha}$ is the onsite energy (which is uniform along the chains but different in each chain), $n_{j, \alpha}$ is the charge density at each site, and $t_{||}^{hop},~ U,~ K$ are the nearest-neighbor hopping, onsite electron-electron, and the nearest-neighbor electron-electron interaction strengths at each chain, respectively, $U^{\prime}$ and $K^{\prime}$ are inter-chain nearest-neighbor and inter-chain next nearest-neighbor interactions, respectively; $t_{\perp}^{hop} (t) = t_{\perp}^{hop} e^{-i\omega t}$ is the oscillatory inter-chain tunneling with $\omega = \frac{2\pi}{T}$, leading to the time-periodicity of the Hamiltonian,  $\mathcal{H}(t) = \mathcal{H}(t+T)$. The oscillatory inter-chain tunneling couples to the charge of the particles only, and will lead to the notion of weak local charge-FDS if and only if $U^{\prime} = U$ and $K^{\prime} = K$. It is in this limit the undriven Hamiltonian (i.e., time independent $\mathcal{H}_{hop}^{\perp}$) exhibits an SU(2) symmetry due to the ladder structure, see Supporting Information (SI) Sec. S1 for a proof. Such a Hamiltonian is directly realizable in a DQ-array setup  \cite{Hensgens2017,Sigillito_PhysRevApplied2019,Mills2019}, and also in optical lattice set-ups \cite{Gross995}. A similar set-up has been realized for 4 coupled QDs \cite{Riggelen_APL}.

 In order to evaluate currents and other dynamical observable, we study the time evolution of the density matrix of the system under Floquet-Lindblad quantum-master equation \cite{ Hartmann_2017,Lindblad, GKS, Breuer, Archak_Dhar_Kulkarni_PhysRevA},
 \begin{eqnarray}\label{Lindblad}
 \frac{d \rho}{dt} = \mathcal{L}_{t} [\rho] = -i [\mathcal{H}(t), \rho] + \sum_{\mu} \left(V_{\mu} \rho V_{\mu}^{\dagger} - \frac{1}{2} \lbrace V_{\mu}^{\dagger} V_{\mu}, \rho \rbrace \right),
 \end{eqnarray}
($\hbar = 1$) with a periodic Liouvillian $\mathcal{L}_{t} = \mathcal{L}_{t+T}$ guaranteed by the periodic Hamiltonian $\mathcal{H}(t) = \mathcal{H}(t+T)$; index $\mu$ representing a summation over all Lindblad operators. \XZZ{Periodic drive can be implemented via time dependent control of tunnel barriers $T_{j}$'s, see Fig. \ref{fig_set-up}(b)  \cite{Mills2019}.} Charge flow is modeled by setting Lindbald operators $V_{in}=\sqrt{\gamma_{L}}c^{\dagger}_{1,b}$ and $V_{out}=\sqrt{\gamma_{R}}c_{N,a}$ which push charges into and out of the ladder, respectively (see Fig.~\ref{fig_set-up}). Dephasing is introduced via Zeno-type dephasing operators $V_{j}=\sqrt{\Gamma} (c^{\dagger}_{j,a} c_{j,a}+ c^{\dagger}_{j,b} c_{j,b})$, which act simultaneously on the $(j,a)-(j,b)$ ladder pairs (see SI Sec. S1-A for a detailed discussion). In a QD array, for example, such a dephasing can be induced via suitable placement of quantum-point contacts,
\cite{Contreras_Pulido_2014, field1993measurements, young2010inelastic, levinson1997dephasing,Granger2012} e.g., in a parallel 2DEG layer below the QD simulator \cite{Bounouh_IEEE, Nield_JAP, MICOLICH2002841}. If the dephasing is applied on each QD separately then it does not satisfy the dynamical symmetry criteria (see SI Sec. S1-A) and would destroy the DTC. 

We aim to show that in a two-leg ladder system DTC can be measured via charge-current. We start by showing that a DTC state indeed exists. In SI Fig. S1, we plot the eigenvalues of the Floquet map (the one period time evolution super-operator),
$\hat{\U}_{F} =\hat{\U} (T) = \T \left[ \exp \left(\int_{0}^{T} \hat{\calL} (t')~dt' \right)\right]$ where $\T$ denotes time-ordering \cite{Hartmann_2017}, for a 6-dot system ($N=3$). The eigenvalues that lie on the complex unit circle confirms the existence of a DTC oscillation irrespective of any initial condition \cite{RieraCampeny2020timecrystallinityin, sarkar2021signatures, Gong_PhysRevLett, Gambetta_PhysRevLett_metaDTC}. This means that any initial random density matrix would bring in the DTC oscillation in the long time limit.
Moreover, there can be many DTC states corresponding to the same DTC eigenvalue, and depending on the initial density matrix the system reaches a particular DTC state \cite{sarkar2021signatures}. We choose $t_{||}^{hop}=  U=K = 20\pi~ \text{MHz} = 0.26~\mu eV$, all being electrically tunable. The frequency of the external drive is $\omega = 20 \pi$ MHz (therefore, a period of $T = 0.1~\mu s$) assuming the typical resolution of nano-seconds in the time-dependent measurements in experiments with QD array. For $t_{||}^{hop}= \frac{4}{3} \omega$ we find DTC and $t_{||}^{hop}= \omega$ we find discrete-time quasi-crystal (DTQC). Alternatively, since $\epsilon_{\alpha}$ can be varied via gate voltage, setting the difference between the onsite energies, $(\epsilon_b - \epsilon_a) = \Delta \epsilon = \frac{7}{3} \omega$ leads to DTC, $\Delta \epsilon = 0$ would lead to DTQC for $t_{||}^{hop}=t_{\perp}^{hop}$. All the conclusions we obtain remain true for DTQC also. The system parameters chosen for the numerical calculations are well within the reach of present day experiments \cite{Barthelemy_2013, Kouwenhoven_2001, Hensgens2017, Mills2019, Zajac439, burkard2021semiconductor, Bhattacharya_nano_lett, Lawrie_nanolett, Jang_nanolett}. 

Having a DTC state is not enough; one must show that this state has a signature on a measurable quantity (e.g., currents). To show this, we proceed in the following way. First, we demonstrate that the DTC behavior is manifested through the inter-chain tunneling currents, which are not directly measurable. Then, we proceed to show how the charge-current through the system relates to the inter-chain current, thus proving that the charge-current exhibit signatures of the DTC state.

\begin{figure}
 \centering
 \includegraphics[keepaspectratio=true,scale=0.35]{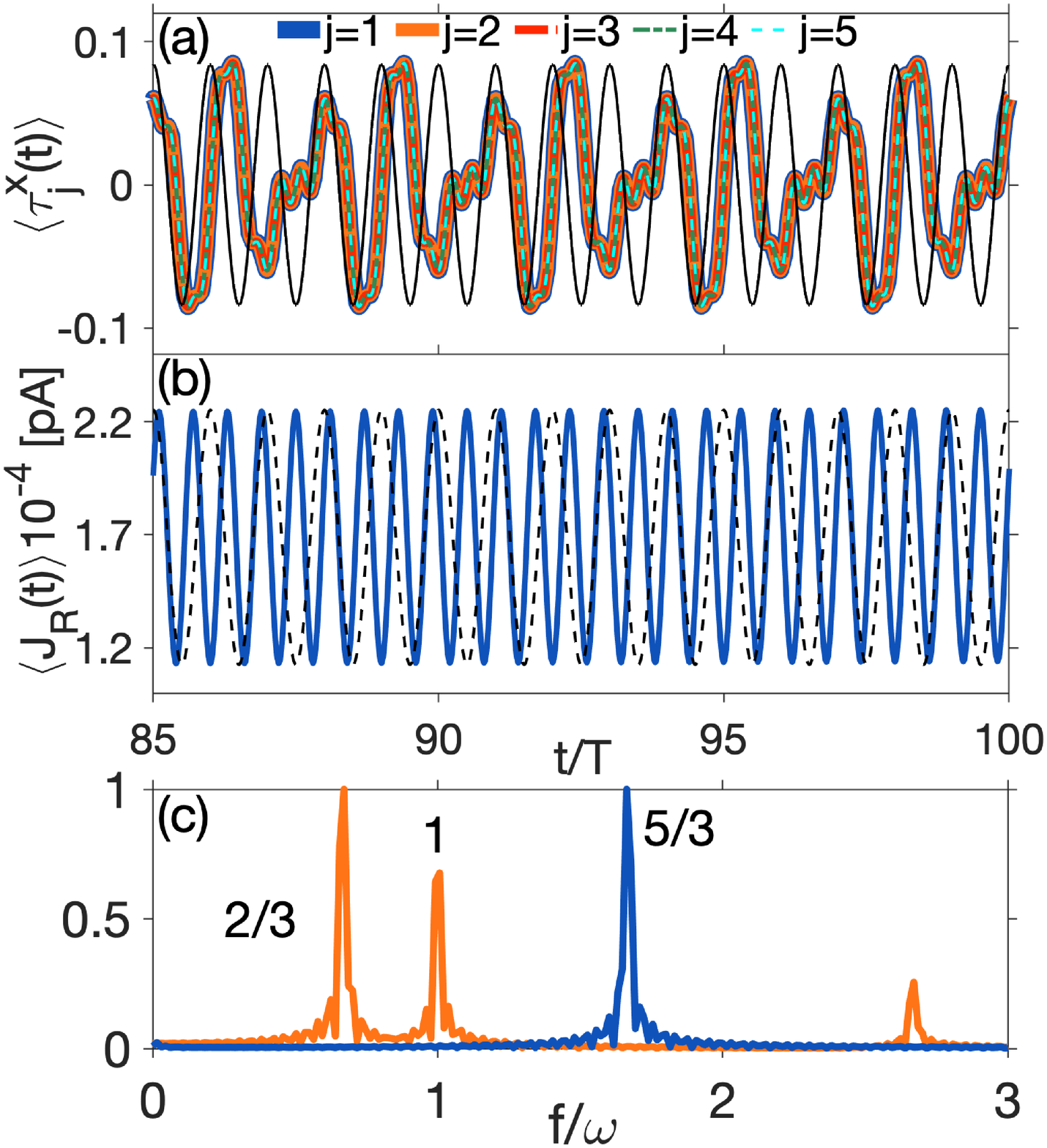}
 \caption{Plot of DTC oscillation and oscillating charge-current. (a) For a system with N=5, plot of $\langle \tau_{j}^{x}(t)\rangle$, and (b) of charge-current $\langle J_{R} (t)\rangle$ through right electrode as a function of time in the long time limit. (c) Discrete Fourier transform $|\tilde{\tau}_{3}^{x}(f)|$ (\textcolor{ppk}{orange line})
 of $\langle \tau_{j}^{x}(t)\rangle$ and $|\tilde{J_{R}}(f)|$ (\textcolor{sph}{blue line}) of $\langle J_{R}(t) \rangle $, respectively. The peak of $|\tilde{J_{R}}(f)|$ at $5/3$ directly provides the value of the Floquet coherence. Value of parameters: $t_{||}^{hop} =  U = K = \omega = 20\pi MHz, ~ t_{\perp}^{hop} = \frac{4}{3} \omega, ~ \gamma_{L} = \gamma_{R} = 10^{-4} \omega$.}
 \label{fig_result_1}
 \end{figure}

In Fig. \ref{fig_result_1} (a) we plot the synchronized oscillation of the real part of the time dependent particle tunneling from the chain $a$ to chain $b$, viz., $\langle \tau_{j}^{x}(t) \rangle$ on the $j'$th inter-chain bond, $\tau^{x}_j=\text{Re}[c^{\dagger}_{j,b}c_{j,a}]$ in the presence of electrodes. We shall show later that the operator $\tau_{j}^{x}$ represents the $x-$component of the effective dipole moment appearing along the inter-chain bonds. We set $\gamma_{L} = \gamma_{R} =\gamma = 10^{-4} \omega$, and choose $\Gamma = t_{hop}$ for faster quantum-synchronization of oscillating effective dipoles residing on the $(j,a)-(j,b)$ bonds (the choice of magnitude of $\Gamma$ only determines the synchronization time scale) \cite{sarkar2021signatures}. In the long time limit between $85T - 100T$, oscillations show clear signature of DTC with period $3T$. This is further fingerprinted in the Discrete Fourier transform (DFT)~ $|\tilde{\tau}_{3}^{x}(f)|$ of the effective-dipole oscillation in the form of a sharp peak at a frequency $f = 2\omega/3$, as seen in Fig. \ref{fig_result_1} (c). For stronger system lead couplings the DTC oscillation decays with a characteristic decay time $\approx 2/\gamma$, see SI Sec. S3 for further details. 

Although the effective dipole $\langle \tau_{j}^{x}(t) \rangle$ shows DTC oscillations, it is not a directly measurable quantity. However, since the system is attached to a pair of electrodes, one can detect the oscillating charge imbalance induced by the effective-dipole oscillation at each inter-chain bond of the ladder through the (charge) current. Fig. \ref{fig_result_1} (b) shows the current $\langle J_{R}(t) \rangle = e \gamma_{R} \text{Tr} [n_{N, a} \rho_{DTC}(t)]$ in picco-Ampere (pA) as a function of time, in the long time limit between $85T - 100T$, where $n_{N, a}$ is the density of spin-less Fermions on the $(N,a)$ site of the ladder attached to the right-electrode (See supplementary Sec. S4 for the derivation of charge-current) and $\rho_{DTC}(t)$ is the time-dependent DTC density matrix which exhibits the DTC time period. DFT of the charge-current indicates that the oscillation exhibits a frequency $f_{cc} = \frac{5}{3} \omega$ (subscript `$cc$' denotes charge-current). Notably, $f = f_{cc}~(\text{mod }\omega) = f_{cc} - \omega$, i.e., charge-current, being the primary observable in transport experiments, provides the signature of DTC. 

 Having obtained DTC in the model \eqref{Eq:ham} we seek to understand why our system exhibits time-crystallinity and charge-current manifests the same, from a mathematical perspective. To this end,
we recall that recently it was shown \cite{Buca_nat_comm2019,diss_Flq_DTC_PRL_Ikeda,Buca_prb_2020, buca2021algebraic, Tindall_2020} that in open spinfull systems, for a DTC to survive a necessary condition is the existence of an FDS, a global operator $A$ that satisfies the following conditions, (i) $\U_{F} (A \rho) = e^{-i\lambda T} A \U_{F}(\rho)$, i.e., $A$ becomes an eigen-operator of $\U_F$ with a characteristic frequency $\lambda$ (in general a function of system parameters) corresponding to the Floquet coherence, and (ii) $[V_{\mu},A] = [V_{\mu}^{\dagger},A]=0$, for all $\mu$. 

We first show that the model \eqref{Eq:ham} indeed exhibits a global FDS operator $\tau^{+} = \sum_{j=1}^{N} \tau_{j}^{+}$ in absence of system-electrode coupling if and only if $U^{\prime} = U$ and $K^{\prime} = K$ (see supplementary Sec. S1 for a proof), where $\tau_{j}^{+} = c_{j,b}^{\dagger} c_{j,a}$. Physically, $\tau^{+}$ represents the total particle tunneling operator from chain $a$ to chain $b$, while $\tau_{j}^{+}$ represents tunneling at each pair of sites. In the Floquet basis (rotated frame) the equation of motion for $\tau^{+} = \sum_{j=1}^{N} \tau_{j}^{+}$ can be shown to be generated by a Floquet Lindbladian $\mathcal{L}_{F}$ whose coherent part is governed by a Floquet Hamiltonian $\mathcal{H}_{F} = \mathcal{H}_{hop}^{||} +\mathcal{H}_{int} + \mathbf{E} \cdot \sum_{j}\pmb{\tau}_{j}$. The Floquet Hamiltonian $\mathcal{H}_{F}$ signifies that in the spin-less model an FDS can exist, which we call a charge-FDS, as a consequence of an effective Stark effect under a DC effective electric field $\mathbf{E} = (t^{hop}_{\perp},0,\Delta \epsilon- \omega)$ in the rotated frame. Here, $\pmb{\tau}_{j} = (\tau_{j}^{x},\tau_{j}^{y},\tau_{j}^{z})$ represents an effective electric dipole operator situated on the $j'$th inter-chain (red) bond, and $\tau_{j}^{z}$ represents a particle imbalance on the $j'$th inter-chain bond, see Fig. \ref{fig_set-up}.

When the system is coupled to electrodes, the FDS condition is no longer valid. Recently, we have shown that even if this is the case, DTC signatures can persist if there exists a weak-local FDS, namely a local operator $A_{loc}$ obeying $\U_{F} (A_{loc}\rho)  =  e^{i \lambda T [1 + i \gamma/\lambda]} A_{loc}\U_{F}(\rho)$, where $[V_{in(out)}, A_{loc} ] \propto \gamma$ makes the FDS weak \cite{sarkar2021signatures}. This leads to the appearance of Floquet coherent states $\rho_{m,n}$. 
The Floquet coherent states are built from the Floquet steady state $\rho_{\text{FSS}}$ as, $\rho_{m,n} = (\tau_{loc, |\mathbf{E}|}^{+})^{m} \rho_{\text{FSS}} (\tau_{loc, |\mathbf{E}|}^{-})^n$ (with integer values of $m,n$), satisfying $\mathcal{U}_{F} (\rho_{m,n}) = e^{i(m-n) |\mathbf{E}| T} e^{-(m+n)\gamma T/2} \rho_{m,n},$
and only a few $m,n$ contribute to the weakly decaying coherent states with decay time $\gamma^{-1}$. For $\gamma T \ll 1$ we find stable DTC oscillation in the long time limit, otherwise a meta-stable DTC emerges \cite{sarkar2021signatures}.
 The characteristic frequency of Floquet coherence is given by $\lambda(\Delta \epsilon,\omega,t_{\perp}^{hop}) = |\mathbf{E}| ~\text{mod} \omega$, where $|\mathbf{E}|= \sqrt{ (\Delta \epsilon- \omega)^2 + (t_{\perp}^{hop})^2}$. When $\frac{f}{\omega} = \frac{\lambda(\Delta \epsilon,\omega,t^{hop}_{\perp})}{\omega} = \frac{q}{p}$ (where $p$, $q$ are integers), DTC emerges with periodicity $pT$, otherwise a discrete time quasi-crystal emerges.  

The charge (weak local) FDS we obtain here exhibits the same mathematical structure as that of the conventional FDS obtained in spinfull models \cite{diss_Flq_DTC_PRL_Ikeda, Buca_nat_comm2019, Buca_prb_2020, Tindall_2020}. It protects the quantum coherence between the different particle imbalance sectors, viz., $\frac{\mathbf{E}\cdot \pmb{\tau}}{|\mathbf{E}|} = \tau_{|\mathbf{E}|}^{z}$, while quantum coherence within each particle imbalance sector is eliminated due to the dephasing, which indicates the importance of simultaneous dephasing of $(j,a)-(j,b)$ pair of sites. In long time limit the system reaches a density matrix of the form $\rho_{DTC}(t) = \rho_{FSS} + \sum_{\substack{m,n \\ m,n\neq0}} \left( c_{m, n} e^{i (m-n) \lambda(\epsilon,\omega, E) t} \rho_{m, n} + h.c \right)e^{-(m+n)\gamma t}$ where $c_{m,n} = |\text{Tr}[\rho_{\text{in}}^{\dagger} \rho_{m,n}]|$ corresponding to integer values of $m,n$, $\rho_{\text{in}}$ being the density matrix of the initial state, are the real coefficients of the superposition, $\rho_{FSS}$ being the Floquet steady-states, and only a few $m,n$ contribute \cite{sarkar2021signatures}. 

  Crucially, for a critical value of system parameter $\Delta\epsilon_{c}$ we find that $\frac{\lambda(\Delta \epsilon_{c},\omega, t_{\perp}^{hop})}{ \omega} = \frac{q}{p} = r$, where $r$ is an integer, see SI Fig. S2. This further points out that an oscillation with periodicity $T$ (corresponding to $p=1$), i.e., a Floquet response, emerges.  A natural question arise, can the DTC state recover even after dynamically reaching a state exhibiting a Floquet response? The answer, quite surprisingly, is positive.
  
  \begin{figure}[ht!]
    \centering
    \includegraphics[keepaspectratio=true,scale=0.30]{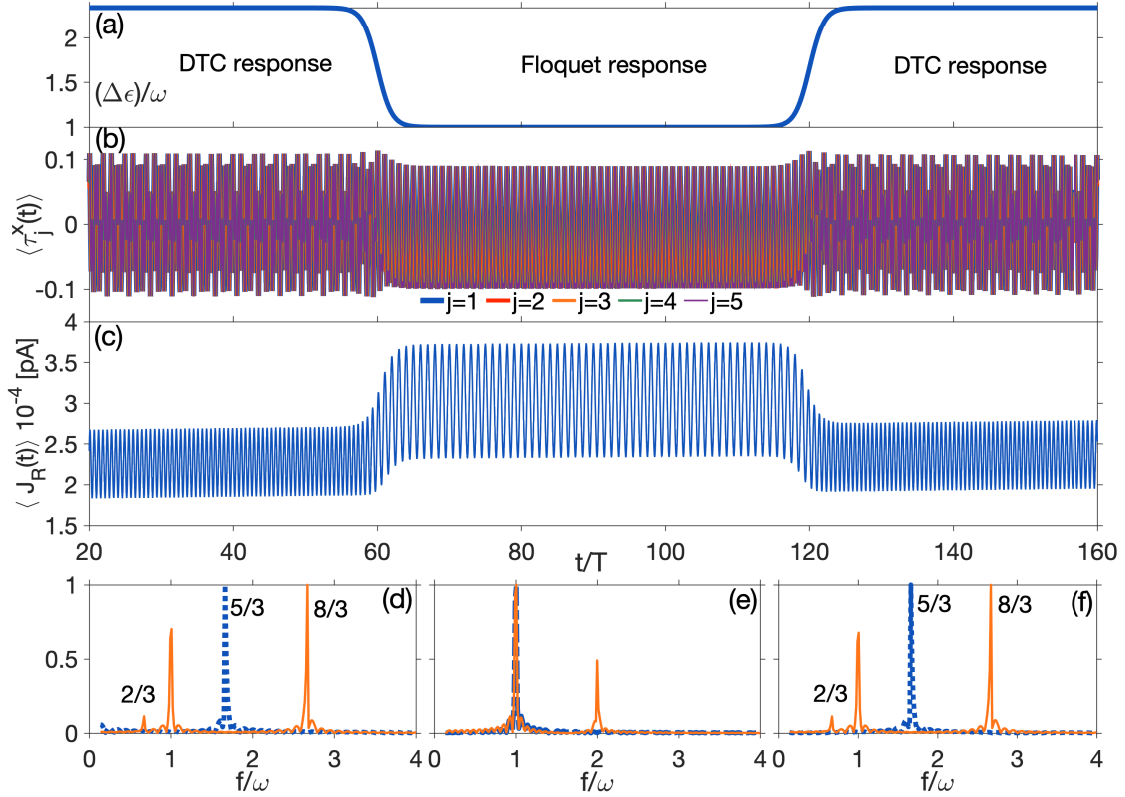}
    \caption{ Dynamical cross-over between DTC and normal phases. (a) Dynamical profile of the onsite energy difference $\Delta \epsilon/ \omega$ as a function of time where $\epsilon_{a}/\omega = 0$.
    For a system with $N=5$, (b) plot of $\langle \text{Re}[\tau_{j}^{+}](t)\rangle$ , and (c) of charge-current $\langle J_{R} (t)\rangle$, in $pA$ through right electrode as a function of time plotted over $180T$. (d), (e), and (f) Discrete Fourier transforms $|\tilde{\tau_{3}^{x}}(f)|$ (\textcolor{ppk}{solid line}) of $\langle Re[\tau_{j}^{+}](t)\rangle$ and $|\tilde{J_{R}}(f)|$ (\textcolor{sph}{dotted line}) of $\langle J_{R}(t) \rangle $, respectively between $5T-55T$, $65T-115T$, and $125T-180T$, respectively. Value of parameters: $t_{||}^{hop} =t_{\perp}^{hop} = U = K = \omega = 20\pi \text{MHz}, ~ \gamma_{L} = \gamma_{R} = 10^{-4} \omega$, strong dephasing strength $\Gamma_{d} = \omega$ ensures fast synchronization. 
    }
    \label{fig_result_3}
   \end{figure}


  To show this, we quench (by ramping) the value of $\Delta \epsilon$ during the dynamics as shown in Fig. \ref{fig_result_3} (a). Fig. \ref{fig_result_3} (b) and (c) plot  $\langle \tau_{j}^{x}(t) \rangle$ and the charge-current $\langle J_{R}(t) \rangle$ from the right lead, respectively, showing that the system dynamically enters, after a transient relaxation, into a normal state exhibiting Floquet response from a DTC state once the system reaches the required critical value $\Delta \epsilon_c$, and again relaxes back to the DTC state once the system is quenched away from $\Delta \epsilon_{c}$. Fig.\ref{fig_result_3} (d), (e), and (f) plot the DFTs of both $\langle \tau_{j}^{x}(t) \rangle$ and $\langle J_{R}(t) \rangle$ showing that the initial DTC response exhibiting a peak at $\frac{|\mathbf{E}|}{\omega} = \frac{5}{3}$ in the DFT of $\langle J_{R}(t) \rangle$, seen in Fig.\ref{fig_result_3} (d), is obtained back, as seen in Fig.\ref{fig_result_3} (f), after the system dynamically passes trough states exhibiting Floquet response seen in Fig.\ref{fig_result_3} (e). In supplementary section S4 we discuss dynamical transition from DTQC to normal phase.

  These results imply that once the system relaxes to a DTC state it can never reach the FSS although it can reach a coherent state exhibiting a Floquet response which is degenerate to the FSS. This shows the stability of the DTC state against continuous change of the system parameters. 

We identify a paradigmatic experimental setting of a QD array placed between leads that can identify and directly measure the appearance of DTC (and DTQC) through the oscillations in the charge-current. 
A charge-FDS emerges as a consequence of an effective `Stark effect' in the rotated frame and protects DTC in the presence of dissipation. This is complementary to the existing notion of the FDS that appear, as a consequence of an effective Zeeman effect, in spin-full Hubbard and XXZ spin models \cite{diss_Flq_DTC_PRL_Ikeda, Buca_nat_comm2019, Buca_prb_2020, buca2021algebraic}.
Furthermore, we show that DTC behavior is surprisingly robust against continuous change of system parameters and support a reversible dynamical transition from DTC state to a state with Floquet response.

Out theoretical results are obtained by keeping a QD simulator in mind where parameters can be controlled via gate voltages \cite{Barthelemy_2013, Kouwenhoven_2001, Hensgens2017, Mills2019, Zajac439}. We choose $\gamma$, the tunneling rate between the source (and drain) and QDs, in the range of $10^{-3} \omega = 62~kHz$ and $10^{-5}\omega = 6.2~kHz$, which is a standard in transport experiments with QD-array set-up \cite{Gustavsson_PhysRevLett}. Although numerical results are obtained in a set-up of 10 QDs, our analytical formulation is valid for any number of QDs. \XZZ{In addition, we have verified numerically that the minimum number of QDs necessary for the FDS based DTC to appear is four (i.e., N=2), a system size that is routinely achieved in experiments, e.g., Refs. [\onlinecite{Hensgens2017, Qiao_PhysRevX_2020}].} The novelty of our finding is that one can detect dissipative-DTC in a quantum transport experiment performed on a QD array and perhaps in other nano-scale systems \cite{Petiziol2021, Hussain_jacs_2018, Chiesa_jpclett2020} for over 100 driving periods via measuring charge current of sizable amplitude. Moreover, our identification of a transport observable of DTC takes the DTC physics beyond the existing cold-atom set-ups and connects DTC to quantum transport phenomena in nano-scale systems.
Although we have used realistic parameters from QD-array set-up \cite{Barthelemy_2013, Kouwenhoven_2001, Hensgens2017, Mills2019, Zajac439}, ultra-cold quantum gases also provide another promising route \cite{Gross995, Brantut1069, Chien2015, Esslinger_annurev-conmatphys}. 

\section{Supporting Information}
\titlecap{charge-FDS of the system, Spectrum of Floquet propagator, Decay of DTC due to system lead coupling, Current operator, Dynamical transition from DTQC to normal state}

\bibliography{tc_bib}

\newpage

\begin{figure}
    \centering
   \includegraphics[keepaspectratio=true,scale=1.0]{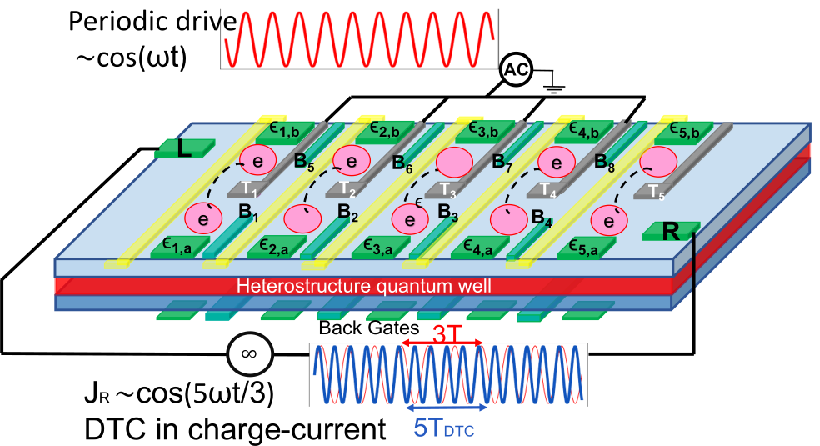}
    \caption{“For Table of Contents Only”}
    \label{For Table of Contents Only}
\end{figure}
\end{document}